\documentclass[letterpaper,twocolumn,10pt]{article}
\usepackage{usenix}

\usepackage{amsmath}

\usepackage{filecontents}
\usepackage{ulem}
\usepackage{hyperref}
\usepackage{pifont}
\usepackage{tikz}
\newcommand*\circled[1]{\tikz[baseline=(char.base)]{\node[shape=circle,fill=black,text=white,inner sep=0pt,minimum size=1em](char) {#1};}}

\newcommand{\sys}{{TraCT}}

\begin{document}
%-------------------------------------------------------------------------------

%don't want date printed
\date{}

\title{{\sys}: Disaggregated LLM Serving with\\CXL Shared Memory KV Cache at Rack-Scale}

\author{
{\rm Dongha\ Yoon}$^{1}$\footnotemark[1],
{\rm Younghoon\ Min}$^{2}$,
{\rm Hoshik\ Kim}$^{2}$,
{\rm Sam H.\ Noh}$^{1}$,
{\rm Jongryool\ Kim}$^{2}$ \\
\\[-4pt]
$^{1}$Virginia Tech \quad
$^{2}$SK Hynix America
}
\maketitle
\footnotetext[1]{This work was done during the first author’s internship at SK hynix.}
%-------------------------------------------------------------------------------
\begin{abstract}
%-------------------------------------------------------------------------------
%%\cdongha{Your abstract text goes here. Just a few facts. Whet our appetites.
%%Not more than 200 words, if possible, and preferably closer to 150.}
%%\cdongha{현재 abstract word count 218입니다.}
Disaggregated LLM serving improves resource efficiency by separating 
the compute-intensive prefill phase from the latency-critical decode phase. 
However, this architecture introduces a fundamental bottleneck: 
key/value (KV) tensors generated during prefill must be transferred to 
decode workers, and existing systems rely on RDMA-based network paths 
for this exchange. 
%%As model sizes and context lengths increase, KV transfer dominates both time-to-first-token (TTFT) and peak throughput, and remains highly sensitive to network contention even when prefix reuse is high.
This paper presents {\sys}, a rack-scale LLM serving system that uses 
CXL shared memory as both a KV-transfer substrate and a rack-wide prefix-aware KV cache.
{\sys} enables GPUs to write and read KV blocks directly through CXL load/store and DMA operations, eliminating the NIC hop that constrains existing disaggregated pipelines. 
However, to realize this design, multiple new challenges such as synchronization, consistency, and data management on non-coherent CXL memory need to be addressed.
{\sys} proposes various software solutions such as the two-tier inter-node synchronization mechanism to address these challenges.
We implement {\sys} on the Dynamo LLM inference framework and show that, across static and synthetic workloads, {\sys} reduces average TTFT by up to 
9.8$\times$, lowers P99 latency by up to 6.2$\times$, and improves 
peak throughput by up to 1.6$\times$ compared to RDMA and DRAM-based 
caching baselines. 
%%These results demonstrate that CXL shared memory is a viable, high-performance communication substrate for rack-scale LLM inference.
\end{abstract}

%-------------------------------------------------------------------------------
\vspace{-0.4cm}    \section{Introduction}
\label{sec:intro}
%-------------------------------------------------------------------------------

%%\cnoh{실험 환경 제약을 감안할 때 Rack-Scale을 강조해 주는 것도 좋을 듯. 제목 등.}

\vspace{-0.3cm}    
Large language models (LLMs) continue to grow in scale and capability, driving rapid deployment across industry and research. 
To improve utilization and reduce cost, modern LLM serving systems are increasingly adopting disaggregated architectures that separate the compute-intensive prefill phase from the latency-critical decode phase. 
Systems such as DistServe~\cite{distserve.osdi24}, Splitwise~\cite{splitwise.isca24}, Preble~\cite{preble.iclr25}, and NVIDIA’s Dynamo~\cite{dynamo.github} show that decoupling these phases enables independent scaling and improved throughput. 
However, this architectural shift introduces a new bottleneck, that is, the movement of key/value (KV) tensors. 
As model sizes and context lengths increase, the volume of KV data exchanged between prefill and decode workers grows to hundreds of megabytes per request, making KV transfer a dominant factor in both request latency and peak system throughput.

Today, disaggregated serving stacks overwhelmingly rely on network-based KV transfers, commonly using RDMA (e.g., UCX, NIXL~\cite{shamis2015ucx, nixl.github}). 
Even when prefix reuse is high, each KV cache hit still requires transporting KV blocks through NIC queues, host DRAM buffers, and layered transport protocols on both ends. 
This network hop significantly inflates prefill latency, increases tail variability, and constrains overall throughput. 
Prefix-aware caching systems such as LMCache~\cite{lmcache.arxiv25} and Mooncake~\cite{mooncake.fast25} improve reuse but still route all KV traffic across the network, leaving network serialization and congestion as persistent performance limitations.

Meanwhile, Compute Express Link (CXL)~\cite{cxl-spec.web} has emerged as a promising substrate for rack-scale systems. 
CXL Type-3 devices provide large, byte-addressable memory pools that multiple hosts may map concurrently using load/store semantics, without involving the network stack. 
This raises a natural question: Can CXL shared memory replace RDMA as the transport substrate for disaggregated LLM serving, eliminating the network hop entirely?

Using CXL to directly publish and consume KV blocks across nodes could fundamentally change the performance and cost profile of LLM inference. Prefill workers could write KV blocks into shared memory via GPU--CXL DMA, and decode workers could read them back with no NIC involvement, no host-to-host copies, and no network-induced variability. However, current CXL devices provide no cross-node atomic operations, do not guarantee coherence across the full device capacity, and expose only raw byte-addressable memory. As a result, even simple operations such as updating metadata, coordinating access, and ensuring visibility require careful software design.

This paper presents \textbf{{\sys}}, a rack-scale LLM serving system that uses CXL shared memory as both (1) a network-free KV-transfer substrate and (2) a rack-wide prefix-aware KV cache. {\sys} tightly couples KV \emph{\textbf{Tra}}nsfer and prefix-aware \emph{\textbf{C}}aching \emph{\textbf{T}}ogether, enabling prefill and decode workers across the rack to communicate through a unified CXL memory pathway. {\sys} removes the RDMA hop entirely: GPUs write missed KV blocks directly into CXL memory and fetch reusable KV blocks back through direct GPU--CXL DMA.

Building {\sys} requires addressing three fundamental challenges imposed by today's CXL hardware:

\begin{enumerate}
    \item \vspace{-0.15cm} \textbf{Inter-node mutual exclusion without hardware atomics.}
    CXL Type-3 devices lack cross-node atomic instructions and do not provide global coherence. {\sys} introduces a two-tier software lock consisting of per-node DRAM-resident local locks and a fixed-size global lock array stored in shared memory, coordinated by a lightweight lock manager. This design bounds contention and supports predictable lock acquisition without per-process lock state.

    \item \vspace{-0.15cm}
\textbf{Correct metadata visibility on non-coherent shared memory.}
    Without cross-node coherence, even mutually exclusive writers may expose stale metadata. {\sys} employs fine-grained cacheline flushing, places metadata in a compact control region, avoids CPU access to large KV payloads, and uses \texttt{clflush} (not \texttt{clflushopt}) to guarantee visibility when publishing shared state.

    \item \vspace{-0.15cm}
\textbf{Shared-memory data structures without shared pointers.}
    Virtual addresses differ across processes and nodes. {\sys} adopts offset-based addressing, a shared-memory allocator combining a global chunk allocator with per-node heaps, and a compact object store that publishes only root metadata (e.g., prefix-cache roots), enabling efficient access without pointer rewriting or complex structural updates.
\end{enumerate}

We implement {\sys} on the Dynamo LLM inference framework~\cite{dynamo.github} and compare with the baseline Dynamo runtime with NIXL/UCX and LMCache using various workloads including synthetic workloads generated using Dynamo’s built-in data generator.
We show that CXL shared memory, through the use of {\sys}, is a viable and effective alternative to RDMA for KV transfer. 
Even without caching, {\sys}, with direct GPU--CXL DMA, reduces prefill latency and sustains throughput comparable to RDMA. With caching enabled, {\sys} improves peak throughput by up to 1.6$\times$, reduces average TTFT by up to 9.8$\times$, and reduces P99 TTFT by up to 6.2$\times$ across realistic workloads. 
We show that {\sys} also improves GPU utilization, increases effective PCIe bandwidth, and lowers power consumption, resulting in more predictable and energy-efficient inference performance.

The remainder of this paper is organized as follows. 
Section~\ref{sec:background} provides background on CXL shared memory, LLM inference, 
and the role of KV management in disaggregated serving systems. 
Section~\ref{sec:design} presents the design of {\sys}, including its 
two-tier synchronization mechanism, metadata visibility strategy, 
shared-memory allocator, and prefix-aware KV cache. 
Section~\ref{sec:implementation} details our implementation atop 
the Dynamo--vLLM runtime and describes how {\sys} enables direct GPU--CXL DMA. 
Section~\ref{sec:evaluation} evaluates {\sys} across microbenchmarks 
and end-to-end inference workloads. 
Section~\ref{sec:related} discusses related work, 
and Section~\ref{sec:conclusion} concludes.

\vspace{-0.4cm}
\section{Background and Motivation}
\label{sec:background}

\vspace{-0.2cm}
\subsection{CXL Shared Memory}
Compute Express Link (CXL) is an interconnect standard that builds on top of PCIe and enables high-bandwidth, low-latency communication between CPUs and devices such as accelerators and memory expanders.
CXL Type-3 devices, in particular, provide host-accessible memory through the CXL.mem protocol that is used as a way to extend DRAM capacity beyond what can be installed directly on the CPU socket.

%%\cnoh{아래 문단은 마치 CXL 활용이 보편화 되어 있는 상황에서 이렇게 사용되고 있는 것처럼 서술되어 있음. 실제 CXL을 도입한 시스템은 매우 한정적이고 (어쩌면 Belusa 하나 아닌가?) 이런 상황에서 아래처럼 사용되는 있는 점이 더 부각되도록 서술할 필요가 있음. $\rightarrow$}
%%\dongha{
While recent hardware prototypes demonstrate the promise of CXL-based memory expansion in a single host, practical deployments of shared CXL Type-3 devices—where multiple hosts attach to the same device—remain extremely limited. 
Most existing systems treat CXL memory as a DRAM extension or a near-memory tier, but only a few research efforts explore its use as rack-scale shared memory \cite{cxlshm.sosp23, tigon.osdi25, cMPI.sc25, beluga.sigmod26}.
%%}

In current systems, a CXL Type-3 device typically exposes its capacity as a DAX-capable memory region that can be mapped into user space as byte-addressable memory.
From the perspective of software running on a single node, this memory behaves similarly to local DRAM: loads and stores are issued through the normal memory hierarchy, and DMA engines can read and write the region directly.
However, when multiple hosts attach to the same CXL device, the memory is not automatically kept coherent across hosts.
Recent work reports that coherence, when present at all, is limited to small device-specific regions and does not scale to the full capacity of devices~\cite{tigon.osdi25, memorysharingwithcxl.arxiv}.
As a result, applications that use CXL as a shared memory pool must explicitly manage synchronization and visibility across nodes~\cite{cxlshm.sosp23, tigon.osdi25}.

%%\cnoh{위와 같은 맥락으로 서술 수정 필요. 현재까지 CXL type-3로 실험한 연구가 있었던가? Beluga도 type-2 아닌가?$\rightarrow$}
%%\cdongha{CXL version이 2.0이고, device type은 3이 맞습니다.}
CXL shared memory offers attractive properties for rack-scale systems.
It enables memory-like communication semantics (load, store, DMA) without involving the network stack, avoids extra buffer copies in NICs, and can be shared by multiple compute nodes or accelerators.
At the same time, it introduces new challenges.
CXL Type-3 devices do not expose cross-node atomic operations, so traditional synchronization primitives cannot be directly extended across hosts.
The lack of full-device coherence means that software must reason about cacheline state, flushing, and ordering.
In {\sys}, we leverage CXL shared memory as both inter-node transfer media and LLM's KV cache at rack-scale, but must address these synchronization and coherence issues to ensure correctness and performance.

\vspace{-0.4cm}    \subsection{LLM Inference}

\begin{figure}[t!]
    \centering
    \includegraphics[width=1\columnwidth]{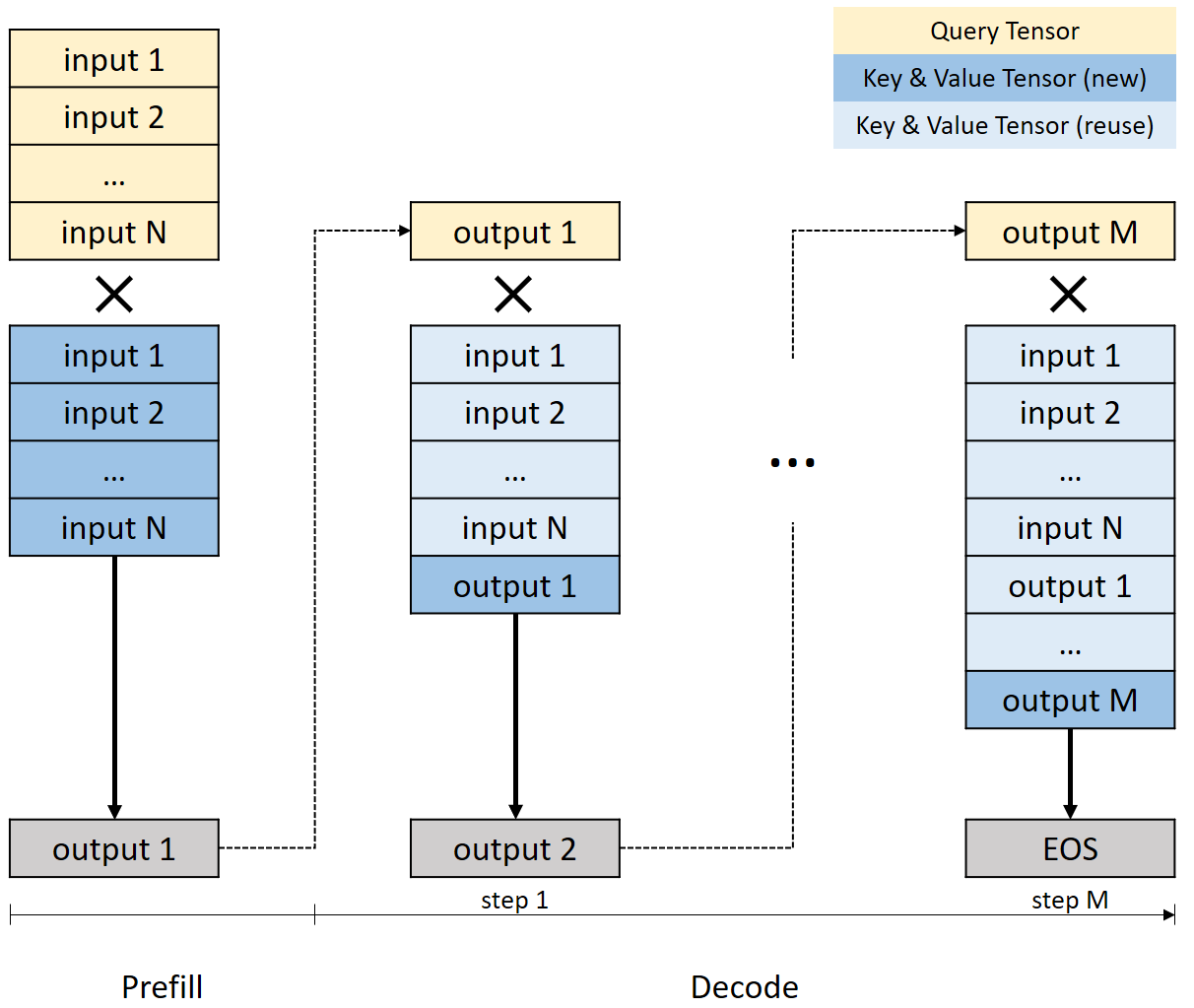}
\vspace{-0.3cm}        \caption{Evolution of KV tensors in LLM inference.}
    \label{fig:bg-inference}
    \vspace{-0.5cm}    
\end{figure}

Modern LLMs such as GPT, Gemini, and Llama~\cite{chatgpt5.web, gemini2.5.arxiv25, llama3.arxiv} are decoder-only transformers that generate tokens autoregressively: each output token is predicted conditioned on all previously produced tokens.  
Inference proceeds in two phases, \textit{prefill} and \textit{decode}, which differ in their compute and memory characteristics.  
Figure~\ref{fig:bg-inference} illustrates how key/value (KV) tensors evolve across these phases.

\medskip
\noindent\textbf{Prefill.}
During prefill, the full input prompt of length $N$ is processed in a single forward pass.
Every layer computes attention across all $N$ tokens, yielding $O(N^2)$ interactions.
This phase is compute-intensive and highly batchable.
The model produces the first output token, and a sequence of $N$ key/value (K/V) tensors—one per input token—which form the initial KV cache.

\medskip
\noindent\textbf{Decode.}
After prefill, inference enters a token-by-token decode phase.
At step $t$, the model computes query (Q)/K/V vectors for the newly generated output token, and attends over all previously stored K/V tensors.
Unlike prefill, per-step computation is small, but the KV cache grows to $N+t$ entries.
Thus, the decode phase becomes memory-bound, not compute-bound.

\medskip
\noindent\textbf{KV cache.}
To avoid re-computing K and V tensors over the entire history at each decode step, inference systems store the intermediate key and value tensors from each layer in a KV cache.
At step $t$, the model reuses the cached K and V from step $1$ to $t-1$ and only computes the Q, K, V for the new token.
For large models and long sequences, the KV cache dominates memory usage; for example, Llama3 405B model, which has 126 layers with 8 heads per layer and a head dimension of 128 requires 504~KB of memory per token.
This makes KV management a first-order concern for both performance and cost of LLM inference systems.

Across recent systems work, the KV cache has emerged as the component that most tightly couples compute throughput, memory capacity, and communication efficiency.
This motivates designs that reduce KV recomputation %%\noh{
to
%%}\dnoh{that is} 
improve reuse or distribute KV storage across multiple memory tiers and nodes
%%\dongha{
~\cite{cacheblend.eurosys25, mooncake.fast25, lmcache.arxiv25, dynamo.github}.
%%}
%%\cnoh{$\leftarrow$ should add references to this line of work}

\subsection{KV Management and Disaggregated LLM serving}
KV caching mechanisms interact closely with how an LLM serving system is architected.
Within a single node, KV tensors are typically kept in GPU memory and managed by the inference engine.
Systems such as vLLM and SGLang, for example, optimize this intra-node KV management.
Specifically, vLLM addresses GPU memory fragmentation and duplication problems by leveraging the operating system's memory paging mechanism, grouping KV tensors into fixed-size blocks, while SGLang introduces RadixAttention, which allows requests with common prefixes to share KV blocks across them~\cite{pagedattention.sosp23, sglang.nips24}.

Beyond a single GPU or node, KV caching has been extended to additional tiers and to multi-node disaggregated settings.
Systems such as Splitwise, DistServe, and Preble partition the pipeline into prefill workers, which run the compute-heavy prompt processing, and decode workers, which perform latency-critical token generation~\cite{splitwise.isca24, distserve.osdi24, preble.iclr25}.
NVIDIA's Dynamo generalizes this disaggregated architecture into a production-level framework~\cite{dynamo.github}.
This separation enables independent scaling of the two phases and improves GPU utilization.
However, every request still requires transferring KV blocks from prefill to decode workers.
Current systems rely on RDMA or similar network paths.
Thus, even when prefix reuse is high, each KV cache hit must traverse the NIC and DRAM on both sides, incurring NIC serialization overhead and extra memory copies, 
%%\noh{
making the approach sensitive
%%}
%%\dnoh{and sensitivity}
to network congestion.

From a KV cache perspective, LMCache and Mooncake demonstrate that KV state can be shared across disaggregated servers.
LMCache provides a KV cache layer that offloads KV blocks from GPU to storage backends, which span from CPU memory to distributed object store, enabling KV reuse across engines~\cite{lmcache.arxiv25}.
Mooncake constructs a distributed KV cache in which CPU memory and SSDs form a global KV cache for separate prefill and decode clusters~\cite{mooncake.fast25}.
However, both systems retain network-based movement of KV tensors for cross-node reuse.

Disaggregated LLM serving therefore improves modularity and resource scaling, but also moves KV transfer into the critical path.
In contrast to network-centric designs, 
% \dnoh{{\sys}}
this study explores using CXL shared memory as a rack-scale KV cache and transfer layer directly accessible to all participating hosts and GPUs through load/store and DMA.
%%\cnoh{\textit{refine $\rightarrow$}
This approach removes the network hop for KV reuse but raises new challenges such as synchronization, consistency, and data management on non-coherent CXL memory.
{\sys} addresses these software support challenges, which we discuss in detail in Section~\ref{sec:design}.
%%but requires software support for synchronization, consistency, and data management on non-coherent CXL memory, challenges addressed in Section~\ref{des:challenge}.
% }

\section{{\sys} Design}
\label{sec:design}
% \dnoh{We propose {\sys}, a rack-scale shared KV cache for disaggregated LLM inference.
% {\sys} exposes a CXL shared memory region that is simultaneously accessible from multiple LLM servers and their GPUs.
% By \textbf{Tr}ansfer \textbf{a}nd \textbf{C}aching \textbf{T}ogether, {\sys} couples KV transfer and prefix-aware caching into a single CXL-based data path, eliminating redundant copies and network hops that dominate in RDMA-based designs.}

% \noh{
\vspace{-0.3cm}    
We propose {\sys}, a CXL-based rack-scale shared KV cache for disaggregated LLM inference.
{\sys} exposes a CXL shared memory region that can be simultaneously accessed by multiple LLM servers and their GPUs.
At its core, {\sys} tightly couples KV \textbf{Tra}nsfer and prefix-aware \textbf{C}aching \textbf{T}ogether within a single CXL-based data path, eliminating the redundant copies and network hops that dominate RDMA-based designs.
% }

\vspace{-0.4cm}    \subsection{Design Goals and Challenges}
\label{des:challenge}
{\sys} is designed with three primary goals.
\begin{enumerate}
\vspace{-0.3cm}    
    \item \textbf{Network-free KV transfer.}
    Replace the RDMA-based network stack with direct DMA between GPUs and CXL shared memory, so that KV blocks flow over the PCIe/CXL fabric instead of NICs.
\vspace{-0.2cm}    
    \item \textbf{Rack-scale KV reuse.}
    Store KV blocks and their prefix-caching metadata directly in CXL shared memory so that any LLM servers in the rack can reuse them without additional copies or network transfers.
\vspace{-0.2cm}    
    \item \textbf{Decentralized KV management.}
    Exploit the shared-memory abstraction of CXL by avoiding centralized metadata servers or coordinator-based KV cache management.  
    Prefill workers and decoding workers should directly read and update shared metadata through load/store operations, requiring no designated owner or manager of KV state.
\end{enumerate}
Achieving these goals introduces the following challenges.

\textbf{(1) Ensuring mutual exclusive access in shared memory region.}
Mutual exclusion is a fundamental requirement in {\sys} because nodes concurrently update shared metadata (e.g., prefix-cache entries, allocator state).
In single-node settings, such metadata is safely protected by hardware cache coherence and atomic instructions.
%%\ddongha{
%%However, current 
% \cnoh{이게 현 제품의 한계가 아니라 fundamental한 문제라는 점을 강조하면 좋을 듯. 이 부분은 섹션 다 읽고 수정 고려해봐라 (뒤에 더 언급하고 있으니).}
%%CXL Type-3 devices expose memory with load/store semantics but without cross-node atomic operations or hardware coherence across the full device capacity.
%%Therefore, standard synchronization primitives such as test-and-set locks cannot be used safely across nodes.}
%%\dongha{
However, this assumption breaks down in multi-host CXL deployments: current CXL Type-3 devices expose memory with load/store semantics but provide no cross-node atomic operations and no full-device hardware coherence.
This limitation is not merely an artifact of first-generation hardware. 
Recent analyses indicate that providing coherence at the scale of multi-terabyte CXL memory is fundamentally impractical due to snoop-filter size, power constraints, and cross-host coherence traffic~\cite{tigon.osdi25, memorysharingwithcxl.arxiv}.  
%%}

Existing CXL shared memory systems take two different approaches to this problem:
(1) Some rely on a small, device-specific coherent region 
% \cnoh{future CXL도 이렇게 밖에 지원이 안될 것이라는 (앞 fundamental 얘기) 점을 언급해 주면 좋을 듯.}
and assume that remote atomics within that region behave correctly~\cite{cxlshm.sosp23, tigon.osdi25}.  
% \ddongha{
This simplifies lock implementation, but the coherent region is typically limited in size and is not guaranteed across devices.
% }
% \dongha{
% While convenient, such coherent zones are limited in size, not architecturally guaranteed, and unlikely to scale in future CXL generations because the underlying coherence limitations remain fundamental rather than merely implementation-specific.
% }
(2) Other systems avoid locks altogether by using producer–consumer queues as the synchronization substrate.
For example, cMPI allocates an $N \times N$ matrix of queues for $N$ participants~\cite{cMPI.sc25}.
This design provides deterministic communication paths but requires $O(N^2)$ memory and forces each participant to scan many queues, increasing CPU cost as $N$ grows.
Beluga reduces this cost by centralizing metadata operations through a single metadata server~\cite{beluga.sigmod26}.
%%\dongha{
This centralized server–client design is common in network-based systems because it simplifies concurrency control and hides coordination complexity behind a single metadata owner.  
However, when applied to CXL shared memory, centralization contradicts the purpose of exposing a load/store–accessible shared region.
Every metadata operation, even simple prefix-cache lookups, must reach the metadata server, reintroducing communication overhead that CXL is meant to eliminate.
In contrast, {\sys} requires that all workers operate directly on shared-memory metadata, without a designated coordinator.
This demands a synchronization mechanism that remains correct in the absence of hardware atomics or coherence.
%%}
% \cdongha{Need to find a point to attack Beluga here.}

% \dnoh{For {\sys}, these limitations mean that a practical mutual-exclusion mechanism must satisfy the following requirements:
% (i) it must operate correctly without relying on hardware atomics or coherence; and
% (ii) it must scale with the number of participating servers without incurring memory or CPU costs.
% This challenge motivates the need for a software synchronization mechanism that uses only load/store accesses to CXL memory while bounding CPU overhead and contention.}

% \noh{
Thus, the challenge for {\sys} is to provide a practical mutual exclusion mechanism that (i) operates correctly without relying on hardware atomics or coherence and (ii) scales with the number of participating servers without incurring memory or CPU costs.
{\sys} addresses this challenge with a novel two-tier software synchronization mechanism that relies solely on load/store accesses to CXL memory while bounding CPU overhead and contention (Section~\ref{sec:mutual-exclusion}).
% }

\textbf{(2) Cache coherence.}
Although the latest CXL specification~\cite{cxl-spec.web} defines snoop-based coherence, recent work shows that coherence---when available at all---is confined to small regions rather than the full device capacity~\cite{tigon.osdi25, memorysharingwithcxl.arxiv}.
% \cnoh{계속 같은 얘기지만, 이게 fundamental한 제약이라는게 중론인데, 이게 대한 근거를 제시하면서, fundamental하다는 점을 앞에서 한 번 언급하면 좋을 듯. 근거를 찾기 어려우면 (또 시간이 없으니), 지금 있는 그대로 가져가도 됨.}
Given terabyte-scale CXL devices, maintaining global snoop filters is impractical~\cite{tigon.osdi25}.
Consequently, even if a software lock ensures mutual exclusion at the abstraction level, concurrent loads and stores to the same cacheline from different nodes can still observe stale data unless caches are explicitly flushed and invalidated.
Naively flushing on every access would quickly exhaust CPU cycles and PCIe bandwidth.
% \cnoh{$\leftarrow$ 이게 Belusa에서 사용하는 방법 같던데. 멘션하는게 좋지 않을까 싶은데.}
% \cdongha{Beluga에서는 clflush 뿐만 아니라 MTTR, intel DSA 등 다양한 방법을 활용하는 것으로 보입니다. (Beluga paper section 5.1)}

% \dnoh{{\sys} must therefore implement fine-grained cache-line management that (i) minimizes the number of flushed lines, (ii) scopes coherence to a small control region, and (iii) preserves correctness for KV metadata and payloads.}

% \noh{
Thus, the challenge for {\sys} is to provide a fine-grained cache-line management mechanims that (i) minimizes the number of flushed lines, (ii) scopes coherence to a small control region, 
% \cnoh{$\leftarrow$ 이 얘기는 작은 영역에 국한시켜도 된다는 얘기 같은데, 이건 이미 제공한다고 봐야 하는게 아닌지? 아니면 다른 얘기를 하려고 한건지?}
% \cdongha{KV tensor데이터는 CXL-GPU direct DMA로 CPU cache를 타지 않기 때문에 메타데이터 coherence만 handle해도 된다는 의미입니다. small region에 대한 coherence를 지원하는 디바이스들이 있지만, 저희 환경에서는 모든 영역이 non-coherent합니다.}
and (iii) preserves correctness for KV metadata and payloads.
{\sys} addresses this challenge by using \texttt{clflush}, an instruction that ensures that the cacheline is evicted from the local cache hierarchy before the instruction completes.
While earlier work claims considerable overhead with this approach~\cite{cMPI.sc25},
% \cnoh{$\leftarrow$ Beluga도 같은 주장하지 않았던가?}
%%\ddongha{~\cite{beluga.sigmod26}}
%%\dongha{
%%~\cite{cMPI.sc25},
%%}
we find this choice to be %%\ddongha{performant}
%%\dongha{
the best for ensuring correctness.
%%}.
% \cnoh{$\leftarrow$ 확인 필요}
% \cdongha{Beluga보다는 cMPI~\cite{cMPI.sc25}를 언급하는게 맞을 것 같습니다. cMPI에서 직접적으로 clflush와 clflushopt의 차이를 잘 설명하고 있습니다.}
We discuss the rationale behind this choice in Section~\ref{sec:cache-coherence}.
% }

%%\cnoh{앞 (1), (2) 나름 개별 논의할 내용이 있다고 봐서 별도 언급하는 것이 좋은데, 아래 3~5는 조금 소소한 문제인 것처럼 느껴지는데, 이들을 합쳐서 (3) Other Challenges 정도의 제목 안에서 서술해 나가면 어떨지? (또는 3번까지 별도로? 3번까지가 library의 주 component 같아 보이네. 그러면 library와 연계해서 그리 서술을 하던가.)}

\textbf{(3) Management of shared objects and memory.}
To share KV blocks and their metadata across nodes, an object abstraction over the CXL region is required.
cMPI, for example, exports a CXL SHM Arena that manages objects via a multi-level hash table~\cite{cMPI.sc25}.
While effective for flat key–value mappings, this design is inefficient for hierarchical structures such as trees or multi-level indices: each element in the tree must be registered as a separate object, stressing the hash tables and complicating key management.
% \noh{
To address this challenge, 
% }
% \dnoh{Instead,}
{\sys} aims to publish only a small set of root objects (e.g., the prefix index) and express the remaining structure as pointer-like links within the shared region.
This requires a clear separation between (i) a shared memory allocator that manages raw space and (ii) a shared object store that enables sharing of objects across nodes.

% \textbf{(4) Limited CXL device bandwidth.}
% \cdongha{이 파트는 김종률 박사님과 미팅 후 넣어 보기로 한 부분인데, 정리가 조금 더 필요할 것 같습니다. half-duplex인 host DRAM과 달리 CXL/PCIe 의 full-duplex를 exploit 한다는 내용입니다.}

% While host DRAM-to-GPU DMA fully exploits PCIe bandwidth, CXL device's bandwidth is capped by the device controller performance.
% However, unlike host DRAM which suffers from bus turnaround
% penalties that reduce their effective bandwidth under mixed
% read-write patterns~\cite{cxlaimbod}

% \textbf{(5) Data structure implementation in shared memory.}
% Virtual addresses are meaningful only within a given process address space.
% When multiple processes on different nodes map the same CXL region, each OS may choose a different virtual base address.
% Thus, a pointer created on one node cannot be dereferenced on another.
% To correctly traverse shared data structures from any node, {\sys} must express all cross-object references as offsets relative to a base of the CXL region and provide a lightweight abstraction for manipulating such offset-based pointers.
%%\cdongha{(4) 제거, (5)는 implementation으로 옮겼습니다.}

\vspace{-0.3cm}    
\subsection{Overview}
\begin{figure}[t!]
    \centering
    \includegraphics[width=1\columnwidth]{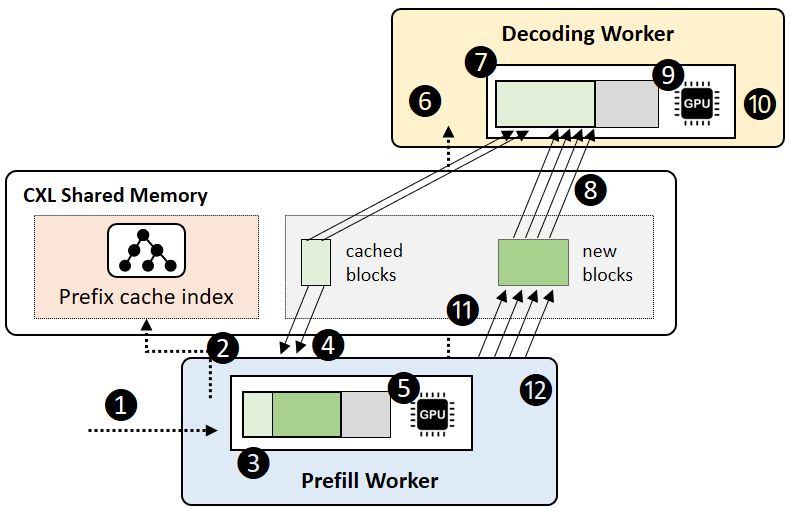}
    \caption{{\sys} overview}
    % \cnoh{library가 맨 아래 있는게 맞나? library는 통상적으로 중간 인터페이스 역할을 한다고 생각한다면 위로 올라가야 할 듯. 새로운 그림을 그린다니, 그냥 참고하시길.}
    % }
    \label{fig:des-overview}
\vspace{-0.5cm}    
\end{figure}

% \cdongha{Need better diagram...}
Figure~\ref{fig:des-overview} illustrates {\sys}'s architecture.
Each rack contains multiple prefill and decoding workers, all connected to a shared CXL Type-3 device.
% \cnoh{아래 있던 문단을 여기로 올렸음. 위치 마음에 안들면 조정하시고.}
{\sys} exposes the CXL device as a byte-addressable, DAX-mapped region on all participating servers.
The region is managed by a library which provides three general primitives: (1) \textbf{an inter-node lock} to ensure mutual exclusion across processes in different nodes (Section~\ref{sec:mutual-exclusion}), (2) \textbf{a memory allocator} for managing limited space (Section~\ref{sec:cache-coherence}), and (3) \textbf{an object store} for metadata sharing across nodes (Section~\ref{sec:shared-object-store}).
Based on the library, {\sys} implements a prefix caching index 
%%\cnoh{$\leftarrow$ prefix caching index가 하나인지, 아니면 여러 개 있는 건지? 여기서는 하나 같은데, 뒤에서는 여러 개라고 되어 있는 듯.}
%%\cdongha{하나입니다.}
and GPU-CXL KV block transfer modules.
Prefill workers insert or update entries in the prefix index, allocate space for KV tensors in CXL memory, and write the tensors via GPU-CXL DMA.
Decoding workers look up prefixes, validate entries, and fetch KV tensors directly from CXL into GPU memory.
%%\cnoh{혹시 시간이 되면, 그림 2에 위에서 언급하는 모듈들도 그림에 추가되면 좋을 듯. 시간 안되면 나중에.}
%%\cdongha{Implementation section의 그림 4에서 언급하고 있습니다.}

%%\cnoh{문단 위치 조정에 따라 아래 문장 추가.}
%%\noh{
The specific steps for inference serving in {\sys} is as follows.
%%}
When an LLM inference request arrives, {\sys} processes the request as follows:
(\circled{1}) \textbf{Prefill Enqueue.} The request is enqueued in the prefill worker's waiting queue.
(\circled{2}) \textbf{Lookup.} Prefill worker looks up the prefix cache for cache-hit blocks.
(\circled{3}) \textbf{Prefill Schedule.} LLM engine 
% \cnoh{vLLM의 특정 시스템인데, 구현과 연계 말고, 일반적으로 표현으로 수정 필요.}
schedules the requests and allocates GPU memory.
(\circled{4}) \textbf{KV Read.} If there is cache-hit, perform CXL-to-GPU KV block transfer.
(\circled{5}) \textbf{Prefill Compute.} Compute cache-missed 
%%\cnoh{$\leftarrow$ missed??}
KV blocks and notify decoding worker to start token generation stage.
(\circled{11}) \textbf{KV Write.} Publish prefix cache entry and copies missed KV blocks (GPU-to-CXL).
(\circled{12}) \textbf{Free.} After KV write, release GPU resources so that next requests can be served.
% \cnoh{번호는 그림에 따라 조정}

% \dnoh{After the \textbf{token generator} receives a request from (5) of KV producer, it processes
% (1) \textbf{Enqueue.} The request is enqueued
% (2) \textbf{Schedule.} Allocate GPU memory.
% (3) \textbf{KV Read.} Read all KV blocks of the prompt. yield GPU cycle until KV blocks are fully loaded to GPU memory.
% (4) \textbf{Compute.} Produce an output token every iteration.
% (5) \textbf{Free.} Release GPU memory.}

% \noh{
% \cnoh{위 문단은 prefill, 아래는 decoding 인듯 한데, 이를 명확히 하고, prefill 부분에서 miss일 때의 작업 서술이 혼란스럽게 느껴짐.}
% \noh{
On the decoding side, when
% }
% \dnoh{When}
a request is received (after (\circled{5})), a decoding worker goes through the following steps:
(\circled{6}) \textbf{Decoding Enqueue.} The request is enqueued
(\circled{7}) \textbf{Decoding Schedule.} Allocate GPU memory.
(\circled{8}) \textbf{KV Read.} Read all KV blocks of the prompt. yield GPU cycle until KV blocks are fully loaded to GPU memory.
(\circled{9}) \textbf{Decoding Compute.} Produce an output token every iteration.
(\circled{10}) \textbf{Free.} Release GPU memory.
% }

\vspace{-0.5cm}    
\subsection{Ensuring mutual exclusive access}
\label{sec:mutual-exclusion}

\begin{figure}[t!]
    \centering
    \includegraphics[width=1\columnwidth]{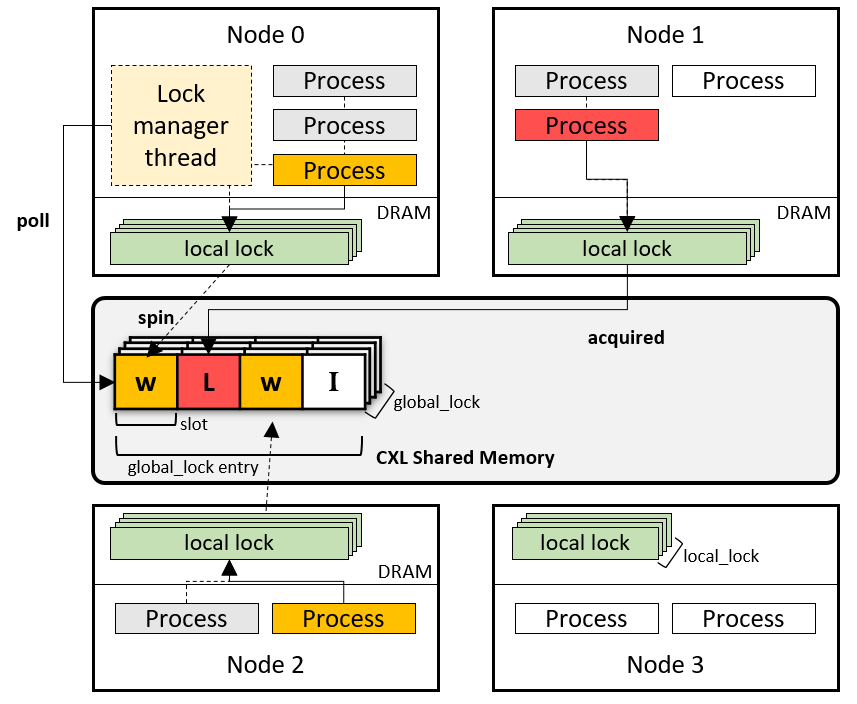}
    \caption{
        % \dongha{
        Two-tier inter-node locking in {\sys}.
        Processes on each node first contends on its local DRAM-resident lock, so that at most one thread per node participates in acquiring the global lock.
        A corresponding paired \texttt{global\_lock}, which is composed of slots, one for each node, resides in CXL memory.
        %%and}
        %%\ddongha{with each entry corresponding to a node} 
        Each slot is in one of three states: \texttt{I} (idle), \texttt{W} (waiting) and \texttt{L} (lock granted).
        One dedicated lock manager thread (residing in Node 0 in this four node figure) guarantees mutual exclusion by granting, at any time, the lock to only a single node.
        % \dnoh{each \texttt{global\_lock} entry has a group of slots corresponds to one node,
        % with \texttt{I} \cnoh{그림에서 .을 I로 수정} (IDLE), \texttt{W} (waiting) and \texttt{L} (lock granted) states.
        % A dedicated lock-manager thread (on a designated node) continuously scans this array and grants the lock to one waiting node at a time.}
        % }
    %%\cnoh{그림에서 중간에 있는 w L w 이런게 뭘 의미하는지? Global lock array 아닌가 싶은데, w, L등이 뭔지? Lock manager thread는 Node 0에만 존재하는 건지?}
    %%\cdongha{w: waiting, L: lock acquried 를 의미합니다. 4칸짜리 블럭 (correspond to 4 Nodes in the example)이 global\_lock 하나를 구성합니다.}
    }
    \label{fig:des-two-tier-lock}
\end{figure}

% \dnoh{To safely share metadata such as allocator state and prefix indices, {\sys} must provide inter-node mutual exclusion without relying on hardware-level atomic instructions, which are unavailable on current CXL Type-3 devices.

% A straightforward design in which every process across all nodes directly participates in a global lock is impractical: processes may appear and terminate dynamically, making the global participant set unbounded; moreover, directly exporting per-process lock state results in excessive contention and forces the lock manager to poll a large and frequently changing set of entries.}

% \noh{
To safely share metadata such as allocator state and prefix indices, {\sys} must provide inter-node mutual exclusion without relying on hardware-level atomic instructions, which current CXL Type-3 devices do not support.
A naïve design in which every process across all nodes directly participates in a single global lock is impractical. 
% \cnoh{차후 얘기지만, impractical하다는 것을 실제로 정량화해서 보여주어야 함. 이를 실험 분석에서 naive 방법도 구현해 보고, 이게 얼마나 영향을 미치는 지 보여주어야 함.
% 그리고, 제안하는 기법이 진짜 얼마까지 scalable한지도 보여주어야 함.}
Processes may join and exit dynamically, making the participant set unbounded, and exposing per-process lock state would create excessive contention while forcing the lock manager to poll a large, continuously changing collection of entries.
% }

% \dnoh{To bound complexity and provide predictable lock acquisition cost, {\sys} adopts a two-tier lock structure consisting of a per-node \textit{local\_lock} and a shared \textit{global\_lock} array, as illustrated in Figure~\ref{fig:des-two-tier-lock}.

% \textbf{Local tier (intra-node exclusion).}
% Each node maintains a conventional lock (e.g. pthread\_mutex), \textit{local\_lock}, in DRAM.
% Only one process per node may attempt to acquire the distributed lock at a time.
% This reduces the number of participants in the global lock from the total number of active processes across the cluster to exactly the number of nodes, which is small, static, and known at initialization.
% This reduction allows {\sys} to maintain a fixed-size global lock array, avoiding per-process metadata and improving scalability.}

% \noh{
To bound complexity and ensure predictable lock-acquisition latency,
% \ddongha{{\sys} adopts a two-tier locking structure consisting of a per-node \textit{local\_lock} and an array of shared\textit{global\_lock}s, as illustrated in Figure~\ref{fig:des-two-tier-lock}.}
% \noh{
{\sys} adopts a two-tier locking structure where a \textit{global\_lock} and \textit{local\_lock} pair work in tandem with both locks being indexed by the same lock identifier.
By acquiring the local lock prior to requesting the global lock, contention for the lock at the global scale can be controlled.
% }

\medskip
\noindent\textbf{Local tier (intra-node exclusion).}
% \ddongha{Each node maintains a conventional DRAM-resident lock (e.g., a \texttt{pthread\_mutex}), which is the \textit{local\_lock}.}
% \noh{
Each node maintains an array of conventional DRAM-resident locks (e.g., \texttt{pthread\_mutex}), which serve as that node’s \textit{local\_lock}s.
Enforcing processes to acquire a \textit{local\_lock} first
% }
% \dnoh{This} 
ensures that at most one process per node attempts to acquire the global lock at any time.
% \noh{
%%\ddongha{This allows {\sys} to use a small, fixed-size global lock array, as the number of nodes is known at initialization, and eliminate per-process metadata leading to significantly enhanced scalability.}
%%\dongha{
% \noh{
This design allows contention for
each \textit{global\_lock} entry to remain a small
% }
% \dnoh{, fixed-size structure}
as the number of participants is bounded by the number of node,
% \dnoh{servers,}
which is known at initialization and is typically limited to a few tens in a rack.  
By collapsing all intra-node contention into the \textit{local\_lock}, {\sys} avoids per-process lock metadata and prevents the lock manager from scanning an unbounded or dynamically changing set of participants.
%%}
% \cnoh{$\leftarrow$ 이런 주장을 뒷받침할 자료가 없어 너무 강조하는 것은 좋지 않아 보임. significant 빼는 것 고려.}
% }
% \dnoh{By reducing global lock participants from all active processes in the cluster to exactly one representative per node, which is small and of a fixed number known at initialization, {\sys} can use a fixed-size global lock array, eliminate per-process metadata, and significantly improve scalability.}
% }

\medskip
\noindent\textbf{Global tier (inter-node arbitration).}
When a process attempts to enter the critical section, as mentioned above, it first acquires its node’s \textit{local\_lock}. 
% \noh{
It then sets its corresponding \textit{global\_lock} pair entry in the CXL shared memory region to \texttt{WAITING} and begins polling that entry. 
% }
% \dnoh{It then sets its node’s \textit{global\_lock} entry in the CXL shared memory region to \texttt{WAITING} and begins polling that entry.}
This polling is a load-based spin on the shared memory word.

% \ddongha{A lightweight \textit{lock\_manager} , which scans the \ddongha{fixed-size}
% \textit{global\_lock} array}
% \ddongha{, runs on one node (Node 0 in our current implementation). scans the fixed-size \textit{global\_lock} array.}
% \noh{
In the background, a \textit{lock\_manager} scans the \textit{global\_lock}s.
For each allocated global lock entry, it probes the per-node slots within that entry and selects one node in the \texttt{WAITING} state to grant the lock by marking its slot \texttt{LOCKED}.
% }
% }
% \ddongha{Its role is limited to selecting one waiting node and marking that entry as \texttt{LOCKED}.}
The manager does not serialize lock usage by holding the lock itself; it merely designates a node that may proceed.
Once the requesting process observes its entry transition to \texttt{LOCKED}, it enters the critical section.

\medskip
\noindent\textbf{Release.}
Upon exiting the critical section, the process resets its \textit{global\_lock} entry to \texttt{IDLE} and releases its \textit{local\_lock}, allowing other processes—both within and across nodes—to compete for future lock acquisitions.

\subsection{Cache coherence}
\label{sec:cache-coherence}
CXL Type-3 devices expose memory with load/store semantics, but do not guarantee coherence across hosts~\cite{tigon.osdi25, memorysharingwithcxl.arxiv}. 
%%\ddongha{
%%beyond small device-specific regions
%%}
Thus, even with software-based mutual exclusion, nodes may observe stale values due to private-cache retention, deferred write-back, or the absence of cross-node invalidation.

% \ddongha{
% This challenge affects two domains in {\sys}:
% (i) correctness of metadata structures such as prefix indices, prefix cache entry, and free lists, and
% (ii) visibility of KV payloads transferred via GPU-CXL DMA.
% To maintain correctness, {\sys} satisfies the following requirements.
% }
% \cnoh{위 문단 내용과 아래 requirements 내용과의 연계를 잘 이해 못하겠음. 아래 내용은 전체적으로 이해 안됨.}

%%\dongha{
To ensure correctness while minimizing performance loss, {\sys} 
% \noh{
takes the following approach.
% }
% \dnoh{choices following methods.}
%%}

\textbf{(1) Metadata visibility.}
% Updates to shared metadata must be explicitly flushed and ordered so that other nodes observe them in program order.
% \ddongha{
% In {\sys}, as metadata is small, they are placed in a compact control region, and as such,
% }
% \dnoh{Because metadata resides in a compact control region,}
%%\dongha{
Updates to shared metadata must be explicitly flushed and ordered so that other nodes observe them in program order.
As metadata structures in {\sys} are compact, {\sys} performs fine-grained cache-line flushing on only the modified lines rather than the entire region.  
This minimizes software coherence overhead while ensuring that every metadata update, such as publishing a KV entry, updating reference counts, or modifying allocator state, is durably written back to the CXL device before other nodes are allowed to consume it.
%%}

%%\cnoh{$\leftarrow$ 이 문장 잘 이해 안감. compact control region이 뭔지?}

\textbf{(2) Payload visibility.}
%%\ddongha{
%%GPU--CXL DMA writes bypass CPU caches.
%%}
{\sys} publishes metadata, such as prefix-cache entries, only after DMA completion, allowing consumers to treat the metadata state as the visibility boundary.
%%\dongha{
Compared to prefix metadata (few cache lines), KV block payloads are large, ranging from hundreds of kilobytes to tens of megabytes depending on the number of tokens per block and model size.
Flushing such regions in software would be prohibitively expensive due to the bulk data movement.  
Fortunately, GPU-CXL direct DMA bypass the CPU caches entirely, so payload data never resides in private CPU caches and requires no explicit flushing.
{\sys} therefore treats the publication of metadata (e.g., setting a prefix-cache entry to \texttt{READY}) as the visibility boundary.
Once metadata is flushed and ordered after DMA completion, all nodes can safely assume that the corresponding KV payload is already durable and globally visible in the CXL device.
%%}
% \cnoh{$\leftarrow$ 미완선 문단}

\textbf{(3) Minimizing cache-line flush overhead.}
KV payloads are never accessed by CPUs during inference and therefore never enter CPU caches.
{\sys} isolates frequently updated metadata into cache-aligned lines while keeping large KV blocks outside the software-coherence critical path.

\textbf{(4) Avoiding incorrect visibility when flushing.}
It is known that \texttt{clflushopt} issues lower-overhead, asynchronous flush requests 
%%\dongha{
compared to \texttt{clflush}~\cite{pmemprogramming.usenix, cMPI.sc25}.
%%} 
However, its behavior is subtle: \texttt{clflushopt} only queues a flush and does not guarantee that the line has reached device memory when the instruction retires.
A subsequent \texttt{mfence} enforces ordering of CPU instructions but does not ensure that pending flushes or stores have propagated to the CXL device.
This can result in incorrect visibility even when mutual exclusion is enforced.
Consider the following simplified code, used to increment a reference count in a prefix cache entry:

\begin{verbatim}
    lock_acquire(entry_lock);
    clflushopt(&ref_count);
    mfence();
    {
        ref_count++;
    }
    clflushopt(&ref_count);
    mfence();
    lock_release(entry_lock);
\end{verbatim}

The reference count prevents {\sys} from evicting prefix-cache entries that are currently in use.
The intended behavior is to flush the old value before the increment and flush the new value afterward.
However, because \texttt{clflushopt} is asynchronous, both flushes may remain pending in the CPU store buffer after \texttt{mfence}.
When the lock is released, another node reading the same cacheline may still observe the old value if the flush has not yet reached the CXL device, violating correctness.

To avoid this situation, {\sys} uses \texttt{clflush}, which ensures that the cacheline is evicted from the local cache hierarchy before the instruction completes.
Although \texttt{clflush} has higher latency, it provides the required visibility guarantees for inter-node correctness when hardware coherence is unavailable
%%\cnoh{$\leftarrow$ clflush clflushopt 이들의 성능의 높고 낮음이 보편적인 지식이 아닐 것이라 생각된다. 그러니 이러한 사실이 설명되어 있는 문헌을 언급하고, 참고 문헌을 추가해야 할 듯.}
%%\cdongha{문단 (4) 앞부분에 reference달았습니다.}

\subsection{Shared Object Store and Memory Allocator}
\label{sec:shared-object-store}
Prefix indices, and other metadata structures must be shared across nodes, but CXL provides only a raw byte-addressable region.
{\sys} therefore designs and implements two foundational mechanisms: a shared memory allocator and a shared object store.

\medskip
\noindent\textbf{Memory allocator.}
%%\cnoh{이 디자인은 lock mechanism과 비슷한 구조. 즉 local global separation. 지금 생각은 이런 맥락의 디자인을 강조해도 좋지 않을까 싶네?  }
%%\noh{
Similar to the two-tier locking structure design,
%%}
{\sys} uses (1) a 
%%\noh{
global
%%} 
\textit{chunk allocator} and (2) per-node local \textit{heap allocator}s.
The \textit{chunk allocator} maintains global bitmap structure in CXL shared memory.
It allocates fixed-size chunks into the \textit{heap allocator} upon request.
The \textit{heap allocator} maintains heap free-lists in local DRAM, enabling cacheline granular memory allocation.
This design eliminates intensive metadata updates in CXL shared memory, shifting metadata contention from inter-node scope to intra-node scope.

\medskip
\noindent\textbf{Shared object store.}
Unlike cMPI's Arena~\cite{cMPI.sc25}, which allocates every element as a separate published object, {\sys} publishes only a few root objects (e.g., the prefix index hash table) and links internal structures via offsets.
This reduces object-management overhead and better supports hierarchical data structures.

\section{Implementation}
\label{sec:implementation}

\begin{figure}[t!]
    \centering
    \includegraphics[width=0.8\columnwidth]{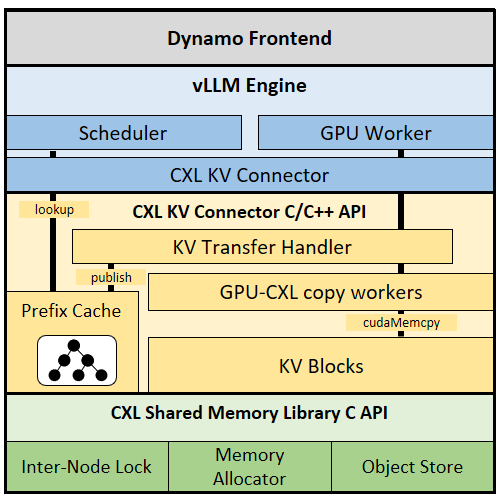}
    \caption{{\sys} software stack}
    \label{fig:des-software-stack}
\end{figure}

Figure~\ref{fig:des-software-stack} illustrates the software stack of {\sys}.
Based on the Dynamo~\cite{dynamo.github} LLM inference framework, {\sys} is implemented as an extension to vLLM’s KV connector layer and a standalone CXL shared-memory library.  
The design allows {\sys} to integrate with Dynamo’s disaggregated prefill/decode pipeline 
%%\dongha{
with only few lines of code modification.
%%}
% without extra modification.
% \cnoh{$\leftarrow$ extra? what do you mean?}

\subsection{CXL Shared Memory Library}
To simplify programming on non-coherent CXL memory, {\sys} provides a lightweight userspace library that exposes a set of C APIs built around the three abstractions described in Section~\ref{sec:design}: locking, shared-memory allocation, and object sharing.  
These APIs hide the low-level details of offset-based addressing, cacheline flushing, and metadata visibility, enabling higher layers of {\sys} (e.g., the CXL KV connector) to operate without directly manipulating CXL-specific primitives.

\noindent\textbf{(1) Shared-memory locks.}  
Applications may allocate and use software-managed locks to protect metadata structures from inter-node contention.
For this, the following interfaces are provided.
\begin{verbatim}
int cxl_shm_allocate_lock(cxl_lock_t *lock);
int cxl_shm_free_lock(cxl_lock_t lock);
int cxl_shm_acquire_lock(cxl_lock_t lock);
int cxl_shm_release_lock(cxl_lock_t lock);
\end{verbatim}

\noindent\textbf{(2) Memory allocation.}  
A malloc-like interface is provided for allocating raw bytes in the CXL region.  
Allocation uses a global chunk allocator combined with per-node local heaps.
\begin{verbatim}
void *shmalloc(size_t size);
void shfree(void *ptr);
\end{verbatim}

\noindent\textbf{(3) Object sharing.}  
A key–reference interface that allows nodes to publish and discover shared objects (e.g., prefix-index roots) are provided.
Keys are stored in a hash table backed by CXL memory, while values are encoded as offsets.
\begin{verbatim}
int cxl_shm_put(char *key, void *ptr);
int cxl_shm_get(char *key, void **ptr);
int cxl_shm_destroy(char *key);
\end{verbatim}

\noindent\textbf{(4) Utility functions.}  
% \noh{
Interfaces to support operations for offset–pointer translation and cacheline eviction are provided.
% }
% \dnoh{Supporting operations for offset–pointer translation and cacheline eviction.}
\begin{verbatim}
shm_ptr_t cxl_shm_get_offset(void *ptr);
void *cxl_shm_get_ptr(shm_ptr_t off);
void clflush(void *addr, size_t size);
\end{verbatim}

Together, these APIs allow higher-level components to safely manipulate shared metadata and KV blocks using only load/store semantics, avoiding the need for kernel modifications or hardware-supported remote atomic operations.

\subsection{Prefix Cache Management}
\noindent\textbf{Block hashing.}
Modern LLM inference systems partition token sequences into fixed-length KV blocks (e.g., 64 tokens per block).
A KV block is uniquely defined not only by its token contents but also by its position in the sequence.
A natural approach for managing these blocks is to maintain a prefix tree.
However, a tree structure would require frequent pointer updates and structural modifications (insertion, split, merge), each of which would incur lock operations and cacheline flushes in CXL shared memory.
Given that {\sys} relies on software-based synchronization on non-coherent memory, such dynamic structures would impose prohibitive overhead.

To avoid these costs, {\sys} uses a fixed-size hash table with linear probing.
Each bucket stores a compact descriptor (hash, pointer to prefix cache entry).
%%\noh{
As the prefix cache size is configured at initialization, the
%%} \dnoh{The} 
hash table is static in size and avoids structural modifications, making it well-suited for a setting where metadata visibility must be ensured through explicit flushing.
%%\cnoh{$\leftarrow$ How come the table size is static? Wouldn't the number of block increase with use?}
%%\cdongha{We configure prefix cache size. hash table is generated at system initialization.}

To generate stable identifiers for KV blocks, {\sys} 
% adopts
%%\dongha{
leverages vLLM’s KV block-hashing mechanism~\cite{vllm.github}.
%%} 
For a block containing list of token IDs \(T_i\), the block hash is computed iteratively:
\vspace{-0.2cm}
\[
    h_i = \mathrm{hash}(h_{i-1}, T_i),
\vspace{-0.05cm}
\]
where \(h_{i-1}\) is the hash of the preceding block.
This construction preserves prefix relationships: identical prefixes produce identical block hashes up to the point of divergence.
Thus, the prefix cache can be indexed without needing to manipulate complex structures.

\medskip
\noindent\textbf{Insertion and lookup.}
When a prefill worker publishes new KV blocks, \textit{KV transfer handler} first performs linear probing on the hash table using the block hash \(h_i\).
After locating an empty bucket, the handler allocates KV storage through the shared-memory allocator and submits a GPU-to-CXL DMA request to the \textit{copy workers}. %%$\leftarrow$ workers? Why copy workers?
%%\cdongha{each worker has its own cudaMemcpy stream. To fully saturate cudaMemcpy bandwidth, multiple workers are needed.}
Once the DMA completes, the handler updates the bucket’s metadata (hash, block length, offset to KV storage) and flushes the corresponding cachelines to ensure global visibility.

Lookup follows the same probing sequence: the consumer searches for \(h_i\) in the table and, upon finding a matching bucket, retrieves the KV offset without modifying any metadata.

\medskip
\noindent\textbf{Eviction.}
{\sys} maintains a simple LRU list in the shared memory.
On every access, the corresponding prefix-cache entry is moved to the end of the list.
When eviction is required, {\sys} selects the oldest entry with a zero reference count, 
% \cnoh{$\leftarrow$ What if there is none with zero reference count? Shouldn't "with lowest reference count" be more appropriate?}
% \cdongha{positive reference count means this entry is in use (need to be pinned until transfer done). Your concern is possible if the CXL cache size is smaller than GPU memory. thus, upcoming requests must wait until victim blocks are available.}
% \cnoh{Exactly. If there is ample cache size, why evict in the first place? Eviction is needed only when there is not enough space.}
marks the entry invalid, frees the associated KV storage, and removes the element from the LRU list.
As eviction only updates compact metadata fields (e.g., reference counts, validity flags, list links) and does not require reorganization of complex data structures, the synchronization overhead remains small.
More sophisticated replacement policies may improve hit rate, but they require substantially richer metadata updates.
% \noh{
Further studies on the tradeoff of replacement policies
% }
% \dnoh{and} 
are left for future work.

\subsection{Data structure implementation in shared memory}
%%\dongha{
Virtual addresses are meaningful only within a given process address space.
When multiple processes on different nodes map the same CXL region, each OS may choose a different virtual base address.
Thus, a pointer created on one node cannot be dereferenced on another.
To ensure correctness across distributed address spaces, {\sys} adopts offset-based addressing for all shared data structures.

\medskip
\noindent\textbf{Offset-based pointers.}
Each shared structure stores 64-bit offsets from the beginning of the CXL region rather than raw virtual addresses.
Each node maintains a local virtual base address for the mapped region, enabling the conversions:
\[
\texttt{ptr} = \texttt{base} + \texttt{off}, \qquad
\texttt{off} = \texttt{ptr} - \texttt{base}.
\]
This ensures that all nodes interpret metadata and payloads consistently regardless of where their OS maps the CXL region in their local address space.

\medskip
\noindent\textbf{Alignment and layout.}
Shared objects are aligned to cache-line boundaries to avoid false sharing.
Frequently written fields are isolated in separate cache lines, while mostly-read fields may be co-located for spatial locality.
{\sys} enforces compile-time layout checks to ensure structure consistency across compilation units and nodes.
%%}

\subsection{Other Details}
\noindent\textbf{Enabling direct GPU--CXL DMA.}
A naive \texttt{cudaMemcpy()} from the CXL-mapped region to GPU memory causes the CUDA driver to allocate an intermediate bounce buffer in host DRAM, introducing an unintended extra copy and reducing throughput.
To avoid this behavior, {\sys} pins the entire CXL shared-memory region using CUDA’s host-memory registration interface.
Once pinned, the CUDA runtime treats the region as page-locked host memory, allowing DMA engines to access the CXL device directly and enabling true GPU-CXL zero-copy transfers.

\medskip
\noindent\textbf{Memory pinning and NUMA placement.}
CXL Type-3 devices attach to a specific CPU socket through that socket’s PCIe root complex.  
As a result, accesses from remote NUMA nodes must traverse an additional inter-socket hop, increasing both latency and bandwidth variability.
To preserve performance, {\sys} binds all threads, including the lock manager and KV connector threads, to the NUMA node directly attached to the CXL device.
This placement minimizes cross-socket traffic, and ensures that CPU-side metadata operations (e.g. prefix cache management, lock acquisition, cacheline flushes) observe consistent and low latency access to the shared memory region.

\medskip
\noindent\textbf{Codebase.}
The full implementation consists of approximately 5K lines of C/C++ for the CXL shared-memory library and the KV connector, plus a small Python wrapper integrating the connector into the Dynamo–vLLM runtime.
% \noh{
We plan to open-source all our sources upon publication.
% }
% \cdongha{Waiting for JR's confirmation}

\section{Evaluation}
\label{sec:evaluation}
% \begin{itemize}
%     \item Request throughput indicate overall system capability.
%     \item TTFT indicates prefill throughput
%     \item ITL (TBT) tells decoding throughput
%     \item tail request latency is critical to user experience
%     \item How to evaluate request fairness? (each req has different prompt length and output length)
% \end{itemize}

% \begin{figure}[t!]
%     \centering
%     \includegraphics[width=1\columnwidth]{v0/F/low_resolution/expr_env.PNG}
%     \caption{Evaluation Environment}
%     \label{fig:eval-env}
% \end{figure}
\begin{table}[t!]
    \centering
     \caption{
        %%\dongha{
            Synthetic workloads generated using Dynamo's request generator.  
            Values denote mean token count, while numbers in parentheses indicate standard deviation.
        %%}
    }
    \label{tab:eval-workload-synth}
    \begin{tabular}{|c|ccc|}
        \hline 
        Workload        &    A  &  B  &  C  \\ \hline
        \hline 
        Input    &               & 4449 (2424) & \\
        \hline 
        Output   &               & 215 (263)   & \\
        \hline 
        Unique   &   1073 (1549)   &   1215 (1693)      & 1631 (2027)\\
        \hline     
    \end{tabular}    
\end{table}

\subsection{Evaluation setup}
% \dnoh{Figure~\ref{fig:eval-env} illustrates our evaluation environment.
% Server~1 runs the benchmark client and all runtime components except KV generation, while server~2 performs KV generation only.}
% \noh{
%%\cdongha{그림이 너무 단순하고, 글로도 충분히 설명이 가능할 것 같아 figure를 없앴습니다.}
%%\ddongha{Figure~\ref{fig:eval-env} depicts our evaluation environment.}
%%\dongha{
Our evaluation environment is configured with 2 servers.
%%}
Server~1 executes the benchmark client and all runtime components aside from prefill worker. Server~2 is provisioned to perform only prefill tasks.
This partitioning is specific to our controlled experimental study in contrast to production systems where these roles are generally co-located within a single node.
% }

% \medskip
\noindent\textbf{Hardware.}
% \dnoh{Each server is equipped with an NVIDIA A6000 GPU (48~GB GDDR), and 512~GB host DRAM.
% For the baseline (NIXL/UCX), the servers are connected with 100~Gbps Mellanox MT2892 NIC.
% For the CXL experiments, we employ Niagara 2.0 device, exposed to the system as a DAX memory region.
% Niagara delivers 640~ns access latency and 10.1~GB/s bandwidth, measured by Intel MLC~\cite{intelmlc.web}.}
% \noh{
Each server is equipped with an NVIDIA A6000 GPU (48~GB GDDR) and 512~GB of host DRAM.
For the baseline (NIXL/UCX~\cite{shamis2015ucx, nixl.github}) configuration, the servers are interconnected using a 100~Gbps Mellanox MT2892 NIC.
%%\cnoh{size of CXL memory? $\rightarrow$}
%%\cdongha{64~GB 입니다.}
For the CXL experiments, we employ a Niagara~2.0 device, a second-generation CXL Type-3 memory expander that provides byte-addressable, load/store–accessible memory through the CXL.mem interface. 
Niagara exposes its capacity to the host as a DAX memory region and functions as a high-capacity, moderate-latency memory tier. 
In our setup, it delivers an access latency of 640~ns and a bandwidth of 10.1~GB/s, measured using Intel's Memory Latency Checker~(MLC)~\cite{intelmlc.web}.
%%\dongha{
We configured 64~GB of space of Niagara for shared memory between workers.
%%}
% }

\medskip
\noindent\textbf{Software.}
% \dnoh{The launch LLM inference runtime using Dynamo v0.5.0, integrated with vLLM v0.10.1.1.
% We compare {\sys} against LMCache~\cite{lmcache.arxiv25}.
% Both {\sys} and LMCache uses 48~GB of prefix cache in CXL shared memory and KV generator's (prefill worker's) host DRAM, respectively.
% To isolate the cost of KV transfer and compare CXL caching ({\sys}) and DRAM caching (LMCache), prefix caching on GPU memory is disabled.
% The model used is DeepSeek-R1-Distill-Llama-8B.}
% \noh{
We deploy the LLM inference runtime using Dynamo v0.5.0 integrated with vLLM v0.10.1.1.
Our evaluation includes three configurations: the baseline Dynamo runtime with NIXL/UCX (denoted NIXL in the results), LMCache~\cite{lmcache.arxiv25} (denoted LMCache), and {\sys} (denoted {\sys}).
LMCache represents a DRAM-resident KV caching baseline, storing a 48~GB prefix cache in the prefill worker’s host DRAM.
{\sys} uses an identically sized prefix cache placed in CXL-attached shared memory to enable pooled, load/store–accessible caching.
The Dynamo/NIXL/UCX configuration serves as a no-prefix-reuse baseline that transfers KV tensors through NIXL over UCX without caching.
To isolate the cost of KV transfer and enable a controlled comparison between CXL-based caching ({\sys}) and DRAM-based caching (LMCache), prefix caching on GPU memory is disabled in all experiments.
All evaluations use the DeepSeek-R1-Distill-Llama-8B model.
% }

%%\subsubsection{Workloads}
%%\label{eval:workloads}

% \dnoh{\noindent\textbf{Static Workloads.}
% These use fixed input/output lengths to directly control the KV transfer size.
% To watch KV transfer effect, output token length is set to 3 and vary the input token length among \{1500, 3000, 4500, 6000\}.

% \noindent\textbf{Synthetic Workloads.}
% Synthetic workloads are generated by Dynamo's data generator.
% Table~\ref{tab:eval-workload-synth} reports the mean (std) of total input, output, and unique length for three workloads.
% The "unique length" affects prefix cache hit rate: larger values imply fewer shared prefixes and thus lower prefix hit rate.}

% \noh{
% \medskip
\noindent\textbf{Workloads.}
We evaluate two classes of workloads, static and synthetic.
The static workloads use fixed input and output lengths to precisely control the volume of KV data transferred.
To isolate the effect of KV transfer, we fix the output length to 3 tokens and vary the input length over {1500, 3000, 4500, 6000}, thereby directly modulating the KV transfer size.

For the synthetic workloads, we generate three request sets using Dynamo’s built-in data generator.
Table~\ref{tab:eval-workload-synth} reports the mean and standard deviation of total input length, output length, and unique length for each workload.
The ``unique length'' captures the diversity of request prefixes and influences the prefix-cache hit rate, where larger unique lengths correspond to fewer shared prefixes and lower expected cache hit rates.
% }

% \cnoh{결과 그림들에서 QPS 값이 1,2,3으로 되어 있는데, 이게 위 실험 환경에서 어떤 경우에 해당되는지 연계가 안됨; CXL-4K, -2K 역시 연계 안됨.}
\subsection{Transfer over CXL}
\begin{figure}[t!]
    \centering
    \includegraphics[width=1\columnwidth]{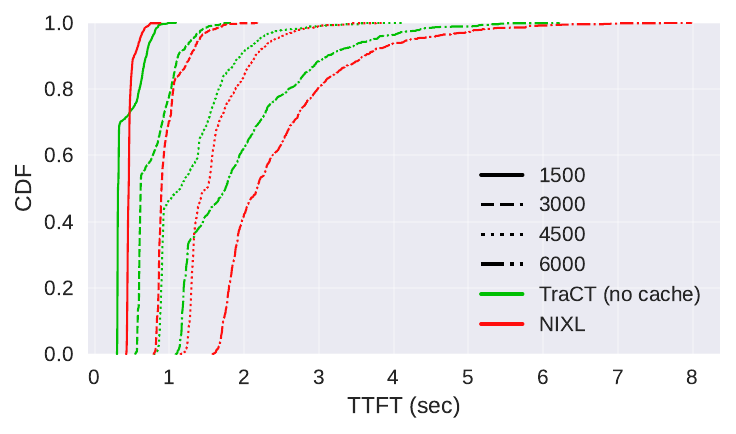}
    \vspace{-0.5cm}
    \caption{{\sys} vs NIXL TTFT CDF. The number indicates input token length.}
    \label{fig:xfer-ttft-cdf}
\vspace{-0.5cm}    
\end{figure}

\begin{figure}[t!]
    \centering
    \includegraphics[width=0.9\columnwidth]{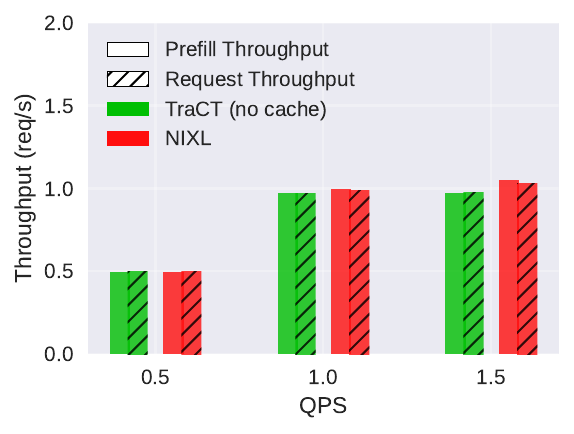}
    \vspace{-0.5cm}
    \caption{{\sys} (no cache) vs.\ NIXL throughput for 6000-token inputs.}
    \label{fig:xfer-throughput}
\end{figure}

%%\dongha{
\medskip
\noindent\textbf{CXL can replace RDMA-based KV transfer.}
Figure~\ref{fig:xfer-ttft-cdf} compares time to first token (TTFT) distributions of {\sys} (with prefix caching disabled) against the NIXL/UCX baseline across four input lengths (1500–6000 tokens).
For all input lengths, {\sys} shifts the CDF curves left, indicating consistently lower prefill latency.
The benefit increases with input size: for shorter prompts (1500 tokens), the TTFT gap is modest, while for the longest prompts (6000 tokens) {\sys} shows a visibly steeper CDF and shorter tail.
Because no prefix caching is used in this experiment, the improvement is attributable entirely to the KV-transfer path—i.e., direct GPU–CXL DMA avoids the NIC queues, host DRAM copies, and transport-layer overhead present in NIXL.

Figure~\ref{fig:xfer-throughput} evaluates both prefill and end-to-end request throughput under increasing QPS for 6000-token inputs.
Across the full load range, {\sys} (without caching) sustains throughput comparable to RDMA-based NIXL, demonstrating that CXL shared memory can serve as an effective bulk KV-transfer path even under high concurrency.

These results show that a CXL-based DMA path provides lower and more stable prefill latency and sustainable throughput than RDMA for KV transfer, even without caching benefits.
Thus, CXL shared memory is a viable transport substrate for rack-scale disaggregated LLM inference and eliminates the network hop that dominates existing prefill–decode pipelines.
%%}
% \subsection{\noh{Performance: Throughput and Latency Performance} \dnoh{KV reuse}}
\subsection{Performance: Throughput and Latency}

\begin{figure}[t!]
    \centering
\vspace{-0.55cm}        \includegraphics[width=0.9\columnwidth]{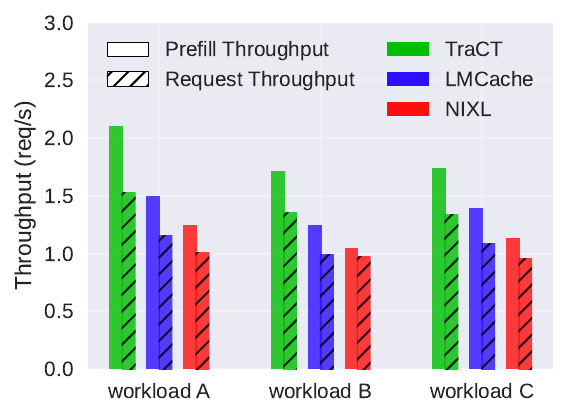}
    \vspace{-0.5cm}
    \caption{Peak throughput.}
    \label{fig:prefix-throughput}
%%\vspace{-0.5cm}    
\end{figure}

%%\cdongha{Note: we will use the term "token generator"/"KV producer" instead of Decoder/Prefill worker.} %%\cnoh{$\leftarrow$ 이런 용어를 써야 하는 이유가 있나? 후자가 더 conventional한게 아닌가?}
%%\cdongha{엄밀히 말하면 Prefill stage는 KV compute 후 first output token generation까지를 의미하는데, 현재 저희가 쓰고 있는 architecture는 prefill에서 KV compute, transfer만 하고 first token은 decoding에서 generation합니다.} \cnoh{ok 일단 그리 둡시다. 나중에 어찌할지 정리.}
%%\cdongha{Decoding / prefill worker를 쓰는게 더 나을 것 같습니다.}
% 
% \dnoh{\textbf{{\sys} significantly improves peak request throughput.}
% Figure~\ref{fig:prefix-throughput} shows overall request throughput and Figure~\ref{fig:prefix-hitrate} reports prefix cache hit rate.
% Although {\sys} shows worse or similar prefix hit rate (\cdongha{\sout{Current eviction policy is naive LRU. cache eviction policy is out of scope of this paper.}}) , {\sys} consistently delivers higher request throughput than LMCache and NIXL.
% With CXL shared-memory caching, KV producer avoids repetitive KV regeneration and KV transmission, allowing decoding workers to immediately fetch reusable KV blocks from the CXL shared-memory, while the KV producer in LMCache must transfer all hit and missed blocks to the token generator.
% As a result, {\sys} achieves up to 1.6× higher peak throughput than LMCache at QPS=3.0, and it maintains stable throughput even as workload pressure increases.
% These gains illustrate that the elimination of extra copy of cache-hit blocks directly translates into higher overall system capacity.}
% \noh{

\begin{figure}[t!]
    \centering
\vspace{-0.8cm}        \includegraphics[width=0.9\columnwidth]{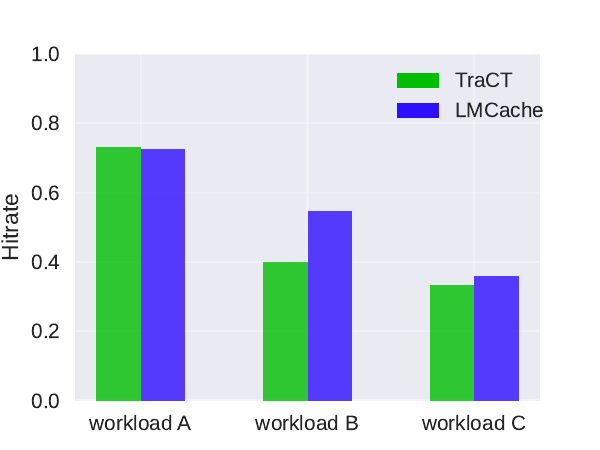}
\vspace{-0.5cm}    \caption{Prefix cache hit rate of workloads.}
    \label{fig:prefix-hitrate}
\end{figure}

\textbf{{\sys} significantly improves peak request throughput.}
Figure~\ref{fig:prefix-throughput} presents the overall request throughput.
Across all load levels, {\sys} consistently delivers higher throughput than both LMCache and NIXL, even though its prefix-cache hit rate is comparable to or even lower than that of LMCache (Figure~\ref{fig:prefix-hitrate}).\footnote{Our current design uses a naive LRU-based eviction policy; exploring more advanced policies is left to future work.}

With CXL shared-memory caching, the prefill worker in {\sys} avoids repeated KV regeneration and network transmission.
Decoding workers can directly fetch reusable KV blocks from the shared-memory pool, whereas LMCache must transmit all blocks, both hits and misses, to the decoding worker.
Consequently, {\sys} achieves up to 1.6$\times$ higher peak throughput than LMCache at QPS=3.0 and sustains stable throughput even under increasing load. These gains demonstrate that eliminating the extra copy and transfer of cache-hit blocks directly translates to higher system capacity.
% }

% \dnoh{\textbf{{\sys} reduces and stabilizes TTFT (prefill latency).}
% The CDF of TTFT (Figure~\ref{fig:prefix-ttft-cdf}) shows that {\sys} shifts the entire latency distribution left relative to LMCache and NIXL.
% At the median, {\sys} improves TTFT by 1.4–1.7×, and at the tail (P99) reduces TTFT by up to 2.1×.
% This improvement arises from two mechanisms: (1) GPU–CXL DMA provides a substantially faster KV transfer path than RDMA. (2) {\sys}'s KV producer does not need to write cache-hit blocks since it is already stored in the the shared memory.
% The steeper CDF curve demonstrates that {\sys} not only accelerates but also stabilizes KV generation under load because it avoids network which is more load-sensitive compared to CXL's PCIe-based physical interconnect.}

\begin{figure*}[t!]
    \centering
    \includegraphics[width=0.9\textwidth]{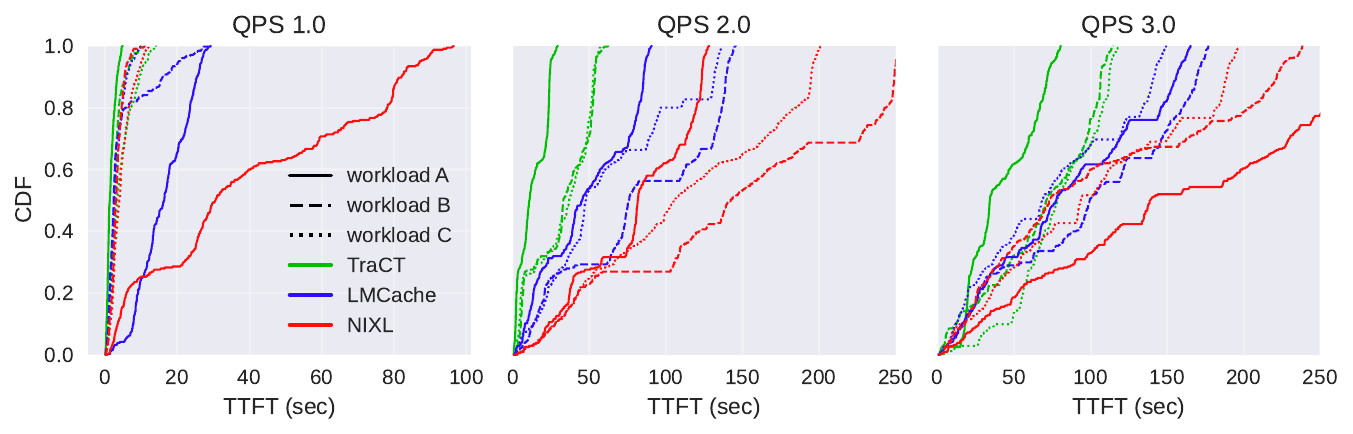}
    \vspace{-0.3cm}
    \caption{TTFT CDF}
    \label{fig:prefix-ttft-cdf}
\vspace{-0.5cm}    
\end{figure*}

% \noh{
\textbf{{\sys} reduces and stabilizes TTFT (prefill latency).}
The TTFT CDF in Figure~\ref{fig:prefix-ttft-cdf} shows a clear leftward shift of the entire latency distribution for {\sys}, indicating uniformly lower prefill latency compared to both LMCache and NIXL.
At the average,
%%\ddongha{median}
%%\dongha{average}
%%,
{\sys} improves TTFT by
%%\ddongha{1.4–1.7}
%%\dongha{
up to $9.83\times$, while at the tail (P99), it is reduced by up to 
%%\ddongha{2.1}
%%\dongha{
$6.2\times$, a critical advantage for maintaining responsiveness under bursty or high-concurrency workloads.

These gains stem from two key design features.
First, GPU–CXL DMA offers a markedly faster and more predictable KV transfer path than RDMA, eliminating the queuing and variability inherent in network fabrics.
Second, {\sys}'s 
%%\ddongha{KV producer}
%%\dongha{
prefill worker
%%} 
avoids rewriting cache-hit KV blocks entirely, as reusable KV data already resides in the shared-memory cache. 
This removes a major source of per-request overhead present in LMCache.

The noticeably steeper CDF curve demonstrates that {\sys} not only reduces latency but it is more stabilized.
By eliminating network involvement and relying on CXL’s PCIe-based local interconnect, which is far less sensitive to contention, {\sys} sustains low variance even under load. 
The result is a tighter, more predictable TTFT distribution and improved end-to-end user-perceived performance.
% }

\begin{figure}[t!]
    \centering
    \includegraphics[width=1\columnwidth]{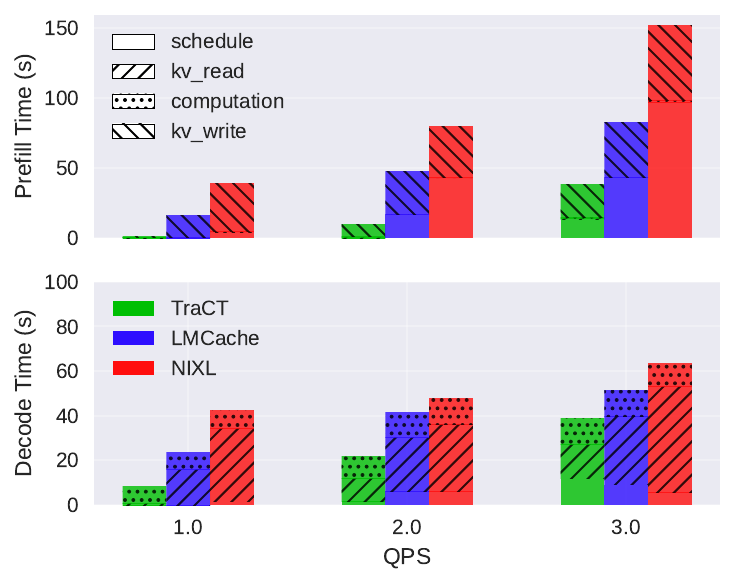}
    \vspace{-0.75cm}
    \caption{Per-request Time Breakdown.
    % \cnoh{시간되면 좀 더 잘 보이게 일부 바를 줌인해 보는 것도 좋을 듯}
    }
    \label{fig:prefix-breakdown}
\vspace{-0.5cm}    
\end{figure}

\begin{figure*}[t!]
    \centering
    \includegraphics[width=1\textwidth]{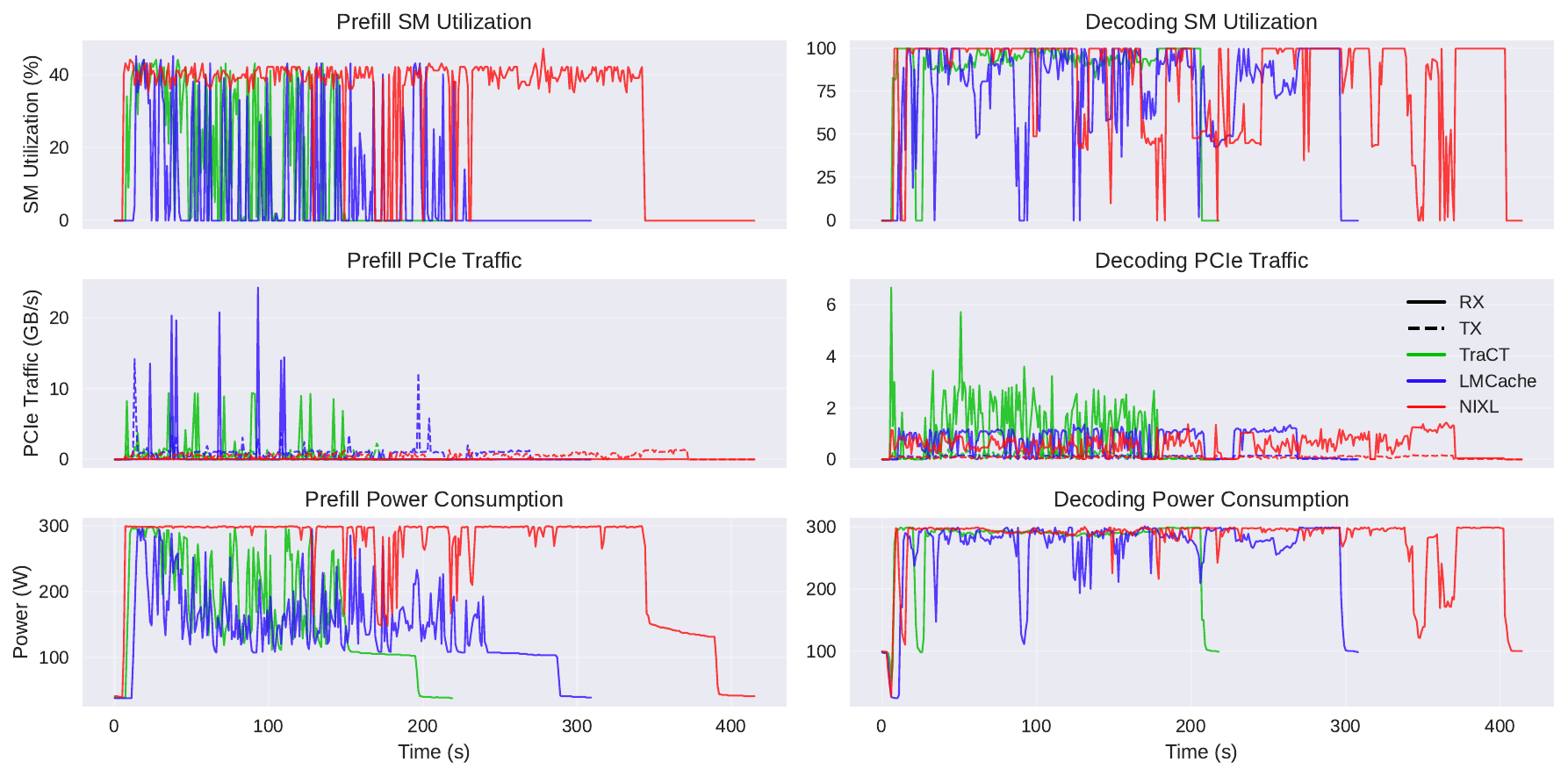}
    \caption{GPU resource consumption over time for prefill and decode workers under QPS\,=\,3.0.
    RX denotes GPU receive bandwidth (PCIe/CXL\,$\rightarrow$\,GPU), and TX denotes GPU transmit bandwidth (GPU\,$\rightarrow$\,PCIe/CXL).
    % \cnoh{RX, TX? 이게 뭔지?}
    % \cdongha{RX는 receive (PCIe to GPU), TX는 transmit (GPU to PCIe)를 의미합니다.}
    }
    \label{fig:prefix-gpu}
\vspace{-0.5cm}    
\end{figure*}

\subsection{Performance Breakdown}
%%\dongha{
\textbf{{\sys} reduces prefill computation and the amount of KV block transfer.}
Figure~\ref{fig:prefix-breakdown} decomposes per-request time into four components: 
\textit{scheduling}, \textit{KV read}, \textit{compute}, and \textit{KV write}.  
\textit{Scheduling} time reflects how long a request waits in the scheduler’s queue before resources become available.
\textit{KV read} time measures the duration of CXL-to-GPU DMA when fetching cached KV blocks.
\textit{Compute} time captures the GPU execution required to generate missing KV blocks.
\textit{KV write} time represents the cost of publishing newly generated KV blocks, either by transferring them to the prefix cache or to the decoding worker, and subsequently freeing GPU memory.

In decoding,
%%}
LMCache and NIXL exhibit growing KV read time under load, reflecting repeated memory copies between GPU, host DRAM, and NIC buffers.
In contrast, {\sys} keeps KV read time nearly constant because KV blocks are directly consumed from CXL memory without network transfers.
%%\cdongha{Need to modify and check plot again. but I expect KV read time of LMCache should be better due to bandwidth gap between DRAM-GPU and CXL-GPU.}
%%\cnoh{$\leftarrow$ 이 부분 확인 후 재작성; 일단 수정 안함}
%%\cdongha{Prefill worker에서 KV read와 compute time이 schedule과 KV write에 비해 상당히 낮아서 그림에는 잘 표시가 되지 않습니다.}
%%\ddongha{
%%LMCache shows better KV read performance since DRAM-GPU copy bandwidth is capped by PCIe while CXL bandwidth is capped by its device performance.
%%However, }
{\sys} shows better overall prefill time as it skips KV transfer for cache-hit blocks thus reduces GPU memory holding time.
Compared to NIXL, both {\sys} and LMCache reduces computation time with prefix hit.
However the impact is negligible since KV transfer dominates overall performance.

% \dnoh{\textbf{{\sys} better utilizes GPU resources with lower power consumption.}
% Figure~\ref{fig:prefix-gpu} shows that {\sys} substantially lowers SM utilization during both prefill and decoding.
% Prefill SM occupancy drops because cache hits bypass KV block computation; decoding SM occupancy stabilizes because KV fetches from CXL memory prevent GPU stalls caused by host–host transfers.
% LMCache shows higher peak RX bandwidth in KV producer side because of PCIe-capped DMA bandwidth.
% However, the remote KV transfer is fundamentally based on RDMA, which limits token generator's GPU RX bandwidth.
% {\sys} provides higher GPU RX bandwidth in token generator. since its token generator directly copies KV blocks from CXL shared memory.
% These observation affects overall GPU power consumption.
% Not only {\sys}'s execution time, but also average power is reduced, providing opportunity for reducing total cost of ownership (TCO) of LLM inference system.}

% \noh{
\textbf{{\sys} better utilizes GPU resources while reducing power consumption.}
% \cnoh{그림 10은 매우 복잡해 보임. 아래 설명과 그림과의 연계가 매우 약함. 어떤 그림의 어떤 부분을 두고 설명을 하는 것인지 알 수가 없음. 그림에서 (가급적 모든 그림들에서) 같은 기법은 같은 색을 쓰도록.}
Figure~\ref{fig:prefix-gpu} shows that {\sys} substantially lowers GPU SM utilization during both prefill and decoding.
Prefill SM occupancy decreases because cache hits bypass KV block regeneration entirely, and decoding SM occupancy becomes more stable because KV blocks are fetched directly from CXL memory rather than through latency-prone host–host transfers.
% \cnoh{RX가 뭔지? 그림에서 TX도 나오던데, 이건 뭔지?$\rightarrow$}
% \cdongha{RX는 receive (GPU read), TX는 transmit (GPU write)를 의미합니다.}
LMCache exhibits higher peak RX bandwidth on the prefill worker side due to PCIe-saturated DMA traffic.
However, its remote KV transfers ultimately rely on RDMA, which constrains the decoding worker’s effective GPU RX bandwidth and introduces contention under load.
In contrast, {\sys} enables the decoding worker to copy KV blocks directly from CXL shared memory, providing consistently higher GPU RX bandwidth and avoiding network-induced stalls.
These effects translate directly into lower GPU power consumption.
As {\sys} shortens execution time and reduces average SM activity, it lowers both instantaneous and overall energy usage, offering a path toward reducing the total cost of ownership (TCO) for LLM inference deployments.
% }

% \section{Related Work}
% \label{sec:related}

% \textbf{Disaggregated inference}
% \ddongha{
% \jr{
% For optimizing the serving of LLMs many LLM servings propose methods for prefill-decode disaggregation.
% Prefill-decode disaggregation, and architectureal approach to optimized LLM inference by running the compute-intensive prefill and memory-intensive decode phases on separate, specialized resources.}
% }
\section{Related Work}
\label{sec:related}
\medskip
\noindent\textbf{Disaggregated LLM Inference}
Recent work on LLM serving increasingly adopts disaggregated architectures that separate the compute-intensive prefill phase from the latency-critical decode phase.
Splitwise generalizes this idea to both homogeneous and heterogeneous device deployments and introduces layer-wise KV-cache transfer to overlap transfer and computation~\cite{splitwise.isca24}.
DistServe proposes an architecture that decouples prefill and decode workers and studies resource provisioning policies for different workers~\cite{distserve.osdi24}.
Preble focuses on KV cache-aware prompt scheduling, dynamically steering requests across prefill and decode instances to maximize utilization and adapt to workload variations~\cite{preble.iclr25}.
NVIDIA’s Dynamo framework brings these ideas into a production setting, providing a general-purpose prefill/decode disaggregation substrate and a KV block manager (KVBM)~\cite{dynamo.github}.

All of these systems treat KV transfer as a network operation: KV blocks flow between prefill and decode workers over RDMA.
Some systems attempt to hide this cost via pipelining or overlapping compute and communication, but the network hop remains in the critical path for each request.
In contrast, {\sys} replaces the network fabric with a CXL Type-3 shared-memory device and performs KV transfer via direct GPU-CXL DMA.
This eliminates host–host copies and network serialization, turning prefill–decode communication into local load/store and DMA operations on a shared address space while retaining the benefits of disaggregated prefill/decode scaling.

\medskip
\noindent\textbf{Multi-tier KV Caching}
LMCache/CacheBlend~\cite{cacheblend.eurosys25, lmcache.arxiv25} and Mooncake~\cite{mooncake.fast25} demonstrate that KV state can be shared across disaggregated servers.
LMCache/CacheBlend generalizes prefix reuse by reusing KV blocks even when prefixes do not align exactly and offloading KV blocks to a variety of storage backends, from host DRAM to distributed object stores, to enable cross-engine reuse.
Mooncake constructs a cluster-wide KV cache in which CPU memory and SSDs form a global KV storage for prefill and decode clusters, and integrates with serving frameworks to exploit reuse across users and workloads.

{\sys} is complementary to this line of work: it adopts a prefix-aware KV cache similar in spirit to LMCache and Mooncake, but places the cache in CXL-attached shared memory that is directly accessible via GPU-CXL DMA.
This allows KV blocks to be served without involving the network at all, changing both the performance envelope (TTFT, tail latency, throughput) of disaggregated KV caching.

\medskip
\noindent\textbf{CXL-based Memory Systems and Shared Memory}
A growing body of work studies how to expose CXL devices as shared memory across nodes.
CXL-SHM presents a general CXL shared-memory substrate and demonstrates its use for distributed data structures and RDMA offload; it assumes device-side compare-and-swap (CAS) support~\cite{cxlshm.sosp23}.
Tigon proposes an in-memory database with CXL shared memory, characterizing coherence behavior and demonstrating that hardware coherence, when available, is limited to small regions and does not scale to the full device capacity~\cite{tigon.osdi25}.
cMPI replaces the network path in MPI communication with CXL shared memory~\cite{cMPI.sc25}.
Beluga explores KV cache management for CXL shared memory and centralizes control-plane operations through a dedicated metadata server~\cite{beluga.sigmod26}.

These systems establish CXL Type-3 devices as practical substrates for shared-memory style communication, but their synchronization designs either rely on device-specific coherent regions or avoid mutual exclusion through queue-based designs.
{\sys} instead targets non-coherent, terabyte-scale CXL memories used as a decentralized KV cache for LLM inference.
It introduces a two-tier locking mechanism, a shared allocator and object store, and a cache-coherence strategy based solely on software-managed cacheline flushing.
To our knowledge, {\sys} is the first system to show that such a design can support rack-scale KV caching and transfer for disaggregated LLM serving on real CXL Type-3 hardware.

% \section{Limitations, Future Work}
\section{Conclusion}
\label{sec:conclusion}

This paper presented {\sys}, a rack-scale LLM serving system that replaces RDMA-based KV transfer with direct GPU–CXL DMA, enabling prefill and decode workers to share KV blocks through CXL shared memory.
To operate correctly on non-coherent CXL Type-3 devices, {\sys} introduces a two-tier inter-node lock, a software-managed data visibility mechanism, and a shared allocator and object store designed for inter-node sharing.
Implemented on the Dynamo–vLLM runtime and evaluated on real CXL hardware, {\sys} improves both TTFT and peak throughput over network-based baselines, demonstrating that CXL shared memory is a practical and high-performance substrate for disaggregated LLM serving.

\newpage
\bibliographystyle{plain}
\bibliography{cite}

%%%%%%%%%%%%%%%%%%%%%%%%%%%%%%%%%%%%%%%%%%%%%%%%%%%%%%%%%%%%%%%%%%%%%%%%%%%%%%%%
\end{document}